\documentclass[prl,aps,twocolumn,superscriptaddress]{revtex4-1}
\usepackage{latexsym}
\usepackage{graphicx}
\usepackage{verbatim}
\usepackage{multirow}

\newcommand{\be}{\begin{equation}}
\newcommand{\ee}{\end{equation}}
\newcommand{\ba}{\begin{eqnarray}}
\newcommand{\ea}{\end{eqnarray}}
\newcommand{\baa}{\begin{eqnarray*}}
\newcommand{\eaa}{\end{eqnarray*}}

\def\be{\begin{equation}}
\def\ee{\end{equation}}
\def\bea{\begin{eqnarray}}
\def\eea{\end{eqnarray}}

\def\C60{A$_x$C$_{60}$}

\def\HgCu3{HgCa$_2$Cu$_3$O$_{8+y}$}
\def\HgCu4{HgBa$_2$Ca$_3$Cu$_4$O$_{10+y}$}
\def\TlCu{Tl$_2$Ba$_2$CuO$_{6+\delta}$}
\def\TlCu3{Tl$_2$Ba$_2$Ca$_2$Cu$_3$O$_{10+y}$}
\def\TlCu4{Tl$_2$Ba$_2$Ca$_3$Cu$_4$O$_{12+y}$}

\def\BiCu3{Bi$_2$Sr$_2$Ca$_{2}$Cu$_3$O$_y$}

\def\8LSCO{La$_{1.88}$Sr$_{.12}$CuO$_4$}
\def\110LNSCO{La$_{1.5}$Nd$_{0.4}$Sr$_{0.1}$CuO$_{4}$}
\def\stage4LCO{La$_{2}$CuO$_{4+\delta}$}
\def\Y248{YBa$_2$Cu$_4$O$_8$}

\def\NbSe2{NbSe$_2$}
\def\TaSe2{TaSe$_2$}
\def\TiSe2{TiSe$_2$}

\begin{document}

\title{ Superconducting Gaps via Raman Scattering in Iron Superconductors }
 \author{Chandan Setty}
\affiliation{Department of Physics and Astronomy, Purdue University, West
Lafayette, Indiana 47907, USA}
\author{Jiangping Hu}
\affiliation{Beijing National
Laboratory for Condensed Matter Physics, Institute of Physics,
Chinese Academy of Sciences, Beijing 100080,
China}
\affiliation{Department of Physics and Astronomy, Purdue University, West
Lafayette, Indiana 47907, USA}
\affiliation{Collaborative Innovation Center of Quantum Matter, Beijing, China}
\begin{abstract}
We investigate non-resonant Raman response for iron-based superconductors using the framework of  an effective $S_4$ model that was recently proposed to capture the essential electronic and magnetic properties of Iron based superconductors. We compute the momentum matrix elements and the resulting Raman vertices exactly for different polarization geometries and amazingly find that   a simple $cos k_x cos k_y$ superconducting gap function is in good agreement with experimental data measured in both iron-pnictides and iron-chalcogenides. The Raman peaks are also matched quantitatively with the measured superconducting gaps by other experimental techniques. The result strengthens the validity of the $S_4$ model and the dominance of the s-wave pairing for iron-based superconductors.
\end{abstract}

\maketitle

The discovery of Iron based superconductors \cite{Hosono2008,Johnston2010} has inspired continued searches for new ways of understanding the mechanism of superconductivity.  The major difficulty in understanding their electronic structure  is the occurrence of multiple bands at the Fermi level.  Although many effective model Hamiltonians - constructed by including different number of orbitals like two\cite{Zhang2008-2BandModel}, three\cite{Wen2008-3BandModel} and five \cite{Kuroki2008-Spm} orbitals - have been proposed, their successes are limited; that is, while simple two and three orbital models leave out several essential features observed in experiments, five orbital models lack analytical and numerical tractability which makes it hard for theoretical calculations to predict meaningful results and therefore, fail to isolate the vital ingredients which control the diverse and interesting phenomena observed in experiments.
\newline
\newline  
Recently it was suggested that the underlying, low energy, electronic structure in the iron based superconductors was governed by an $S_4$ symmetric, two weakly coupled, single orbital models \cite{Hu2012-S4}. The motivation for  such a model, henceforth referred to as  the $S_4$ model, was derived from the fact that if the hopping parameters in the original five band model \cite{Kuroki2008-Spm} are tranformed to a different gauge setting, then, one can use  the effective $d_{xz}$ and $d_{yz}$ orbitals to construct a model that accurately describes the essential band structure near Fermi levels. In the real space picture, the two orbitals can be naturally segmented into two groups : one, coupled to the $As$ atoms on the top of the $Fe$ layer and the other, to those at the bottom, with each group containing contributions from both the sub-lattices. It was also demonstrated that the gauge transformation maps the $s-$wave pairing symmetry in the original one $Fe$ Brillouin Zone (BZ) to the $d-$ wave pairing symmetry in the folded BZ defined on the two sublattices, thus establishing that the relevant symmetry which unifies the different families of high $T_c$ superconductors is a combination of both the hopping and the pairing symmetries. It remains, now, to compute the consequences and predictions of such a model, examine its accuracy, and determine its suitability in describing the observed experimental phenomena.
It is important to note that the critical difference between the $S_4$ model and the previous 2-orbital model\cite{Zhang2008-2BandModel} is that the latter fails in capturing the band dispersions of the hole pockets at $\Gamma$ and the electron pockets at $M$. 
\newline
\newline
Electronic Raman scattering in superconductors \cite{Devereaux1995-CupratesTheory,Dierker1984-RamanTheory,Devereaux2007-RMP} has proved to be a powerful tool for determining the structure and symmetry of the superconducting gap. The Raman scattering vertex, in general, can have contributions from all the irreducible representations (irreps) of the point group of the crystal; however, a proper choice of the polarization geometries of the incoming and outgoing light beams can single out contributions from a single irrep to the total scattering cross section. As a result of the weighted BZ averaging, the quasiparticle energy gaps are probed only in selected directions in the BZ depending on the symmetry of chosen irrep   and can, therefore, give information about the different types of anisotropies in the superconducting gap.  The spectral features in different scattering geometries, pairing symmetries, peak positions, anisotropies, low frequency and temperature behavior, coulomb screening effects and final state interactions have all been taken into account in evaluating the response functions for both the cuprates \cite{Devereaux1995-CupratesTheory,Devereaux2007-RMP} and the pnictides \cite{Chubukov2009-Resonance,Mazin2010-Raman,Boyd2010}. 
\newline
\newline
In this  letter, we evaluate the non-resonant Raman response of the hole/electron doped 122 pnictide and electron overdoped 122 chalcogenide by computing the momentum matrix elements and the resulting Raman vertices exactly (within the tight binding approximation) for different polarization geometries. Here, we amazingly find that using a simple $cos k_x cos k_y$ form of the gap function with a combination of the vertices and certain other properties of the $S_4$ model, one can successfully account for the Raman data reported in \cite{Kretzschmar2013,Muschler2009}.  In this case, the Raman peaks are even consistently matched  with the quantitative values of the superconducting gaps measured by  Angle Resolved Photoemission Spectra (ARPES) experiments in these materials. The result strengthens the validility of this simplest model and the dominance of the s-wave pairing in both iron-pnictides and iron-chalcogenides.

The scattered light intensity in a Raman experiment is written in terms of the differential photon scattering cross section as
\begin{eqnarray}
\frac{\partial^2 \sigma}{\partial \omega \partial \Omega}&=&\frac{\omega_S}{\omega_I}r_0^2 S_{\gamma \gamma}(\textbf{q},\omega)\\
S_{\gamma \gamma}(\textbf{q},\omega) &=& -\frac{1}{\pi}[1+ n(\omega)] Im \chi_{\gamma \gamma}(\textbf{q},\omega), \label{eq:CrossSection}
\end{eqnarray}
where $\omega_I$ and $\omega_S$ are the frequencies of the incident and scattered photons respectively and $r_0 = \frac{e^2}{m c^2}$  is the Thompson radius. The imaginary part of the Raman response function is related to the generalized structure factor $S_{\gamma \gamma}$ through the fluctuation-dissipation theorem (eq. \ref{eq:CrossSection}) and $n(\omega)$ is the Bose-Einstein distribution function. The long wavelength Raman reponse measures effective `anisotropic' density fluctuations 
\begin{equation}
\chi_{\gamma \gamma}(\omega) = \int_{0}^{\beta} d\tau e^{- i \omega_m \tau }\langle T_{\tau} \tilde{\rho}_{\gamma}(\tau),\tilde{\rho}_{\gamma}(0) \rangle \mid_{i \omega_m \rightarrow \omega+ i \delta}
\end{equation}
with
\begin{equation}
\tilde{\rho_{\gamma}} = \sum_{\textbf{k},\sigma} \sum_{n,m} \gamma_{n,m}(\textbf{k}) c_{n,\sigma}^{\dagger}(\textbf{k}) c_{m,\sigma}(\textbf{k}), 
\end{equation}
where $n,m$ denote band indices. If we consider only low energy response (order of $meV$), the interband transitions can be neglected and we have the vertex of the form
\begin{eqnarray*}
\gamma_{n}(\textbf{k}) &=& \hat e^i \hat e^s  + \frac{1}{m}\sum_{j \neq n}\frac{ \langle n, \textbf{k} \mid \hat e ^s p \mid j, \textbf{k} \rangle \langle j, \textbf{k} \mid \hat e^i p \mid n, \textbf{k} \rangle}{\epsilon_{n}(\textbf{k}) - \epsilon_j(\textbf{k}) + \omega^i}\\
&&+\frac{ \langle n, \textbf{k} \mid \hat e ^i p \mid j, \textbf{k} \rangle \langle j, \textbf{k} \mid \hat e^s p \mid n, \textbf{k} \rangle}{\epsilon_{n}(\textbf{k}) - \epsilon_j(\textbf{k}) - \omega^s}.
\end{eqnarray*}
 Here $\hat{e}^{I,S}$ denote the polarization directions of the incident and scattered light respectively.  We  evaluate the vertex exactly within the tight binding approximation \cite{Mohan1993,Vogl1995, Pederson2001,Foreman2002, Cruz1999} and do not resort to the effective mass approximation - one that is valid only when there is a single (or a few) band(s) crossing the fermi level and all other bands much higher than typical incoming light frequency - which is widely used in Raman calculations. Such an approximation completely fails in the $Fe$ superconductors where the spacing between the bands is comparable to the Raman frequency.  The vertex function on the $n$th band can be broken down into various contributions from the irreps of the point group as
\begin{equation}
\gamma_{n}(\textbf{k}) = \sum_{\mu} \gamma_n^{\mu} \Phi_n^{\mu}(\textbf{k})
\end{equation}  
where the index $\mu$ denotes the contributions from the different point group irreps and the functions $\Phi^{\mu}_n(\textbf{k})$ are the corresponding basis functions of the $\mu$'th irrep in the $n$'th band. Screening by the long range Coulomb interaction can be taken into account \cite{Dierker1984-RamanTheory} by including the couplings of the Raman charge density $\tilde{\rho}$ to the isotropic density $\rho$ fluctuations. The screened response is given by
\begin{eqnarray}
\chi_{\rho,\tilde{\rho}}^{s}(\omega)&=& \chi_{\tilde{\rho},\tilde{\rho}}(\omega) - \frac{\chi_{\tilde{\rho},\rho}(\omega)\chi_{\rho,\tilde{\rho}}(\omega)}{ \chi_{\rho,\rho}(\omega)}\label{eq:Response}
\end{eqnarray}
with 
\begin{equation}
\chi_{\rho,\tilde{\rho}}(\omega) = \chi_{\tilde{\rho},\rho}(\omega) = \sum_{\textbf{k}}\gamma_n(\textbf{k}) \lambda_n(\textbf{k},i \omega)
\end{equation}
and
\begin{equation}
\chi_{\rho,\rho} = \sum_{\textbf{k}} \lambda_n(\textbf{k},i \omega)
\end{equation}
where $\lambda_n(\textbf{k},i \omega)$ is the Tsuneto function for the $n$th band defined by(at $T=0$),
\begin{equation}
\lambda_n(\textbf{k},i \omega) = \frac{\Delta_n(\textbf{k})^2}{E_n(\textbf{k})^2} \left( \frac{1}{2 E_n(\textbf{k}) + i \omega}+\frac{1}{2 E_n(\textbf{k}) - i \omega}\right)
\end{equation}
Here $E_n(\textbf{k})^2 = \epsilon_n(\textbf{k})^2 + \Delta_n(\textbf{k})^2$. In Eq. \ref{eq:Response}, the second term is the contribution from the long range coulomb interaction and the first is the bare Raman response. In the case of two bands, the total response can be written as a sum of screened responses for the individual bands plus a mixing term  \cite{Dierker1984-RamanTheory} which vanishes in the special case where the vertices and the gaps are identical for the two bands. The overall Raman response can then be written as
\begin{equation}
Im \chi^{sc}(\omega) = Im \chi_1(\omega) + Im \chi_2(\omega) + Im \Delta \chi(\omega)\label{eq:ImResponse}
\end{equation}
where
\begin{eqnarray*}
\chi_{1,2}(\omega) &=& \sum_{\textbf{k}} \gamma_{1,2}^2(\textbf{k}) \lambda_{1,2}(\textbf{k},i\omega) \\
&&-\frac{\left(\sum_{\textbf{k}} \gamma_{1,2}(\textbf{k}) \lambda_{1,2}(\textbf{k},i\omega)\right)^2}{\sum_{\textbf{k}} \lambda_{1,2}(\textbf{k},i\omega)}
\end{eqnarray*}
and 
\begin{eqnarray*}
\Delta \chi(\omega)&=&  \frac{\sum_{\textbf{k}} \lambda_1(\textbf{k},i\omega)\sum_{\textbf{k}} \lambda_2(\textbf{k},i\omega)}{\sum_{\textbf{k}}\left( \lambda_1(\textbf{k},i\omega)+ \lambda_2(\textbf{k},i\omega)\right)}\\
&& \times \left(\frac{\sum_{\textbf{k}} \gamma_1(\textbf{k})\lambda_1(\textbf{k},i\omega)}{\sum_{\textbf{k}} \lambda_1(\textbf{k},i\omega)} - \frac{\sum_{\textbf{k}} \gamma_2(\textbf{k})\lambda_2(\textbf{k}, i\omega)}{\sum_{\textbf{k}} \lambda_2(\textbf{k},i\omega)}\right)^2
\end{eqnarray*}
We will use the expression in Eq.\ref{eq:ImResponse} to evaluate the response functions in the following derivation.

 We  make a quick recap of the $S_4$ symmetric model described by the authors in \cite{Hu2012-S4}. Our starting point is the following Hamiltonian similar to \cite{Zhang2008-2BandModel} on a single copy containing the $x'z$ on the $A$ sublattice and $y'z$ ($x'$ and $y'$ are along the diagonals to the $Fe-Fe$ bonds) on the $B$ sublattice coupled to each other through the $As$ atoms in between, as
\begin{equation}
H =  \sum_{\textbf{k},\sigma}\psi_{\textbf{k},\sigma}^{\dagger} \left(\epsilon_+(\textbf{k}) - \mu)1 + \epsilon_-(\textbf{k})\tau_3 + \epsilon_{xy}(\textbf{k}) \tau_1\right)\psi_{\textbf{k},\sigma}
\end{equation}
with $\tau_i$ as the Pauli matrices and  $\psi_{\textbf{k},\sigma}^{\dagger} = (c_{1,k}^{\dagger}, c_{2,k}^{\dagger})$ with $c_{1,k}^{\dagger}$ and $c_{2,k}^{\dagger}$ being the electron creation operators at the sublattice sites $A$ ($x'z$ orbital) and $B$ ($y'z$ orbital) respectively.  The band parameters are
\begin{eqnarray*}
\epsilon_{\pm}(\textbf{k}) &=& \frac{\epsilon_x(\textbf{k}) \pm \epsilon_y(\textbf{k})}{2}\\
\epsilon_x(\textbf{k}) &=& 4 t_s cos k_x cos k_y - 4 t_d sin k_x sin k_y \\
&&+ 2 t_{3s} ( cos 2 k_x + cos 2 k_y) \\
&&+ 2 t_{3d} (cos 2 k_x - cos 2 k_y)\\
\epsilon_y(\textbf{k}) &=& 4 t_s cos k_x cos k_y + 4 t_d sin k_x sin k_y \\
&&+  2 t_{3s} ( cos 2 k_x + cos 2 k_y) \\
&&+ 2 t_{3d} (cos 2 k_x - cos 2 k_y) \\
\epsilon_{xy}(\textbf{k})&=& 2 t_1 (cos k_x + cos k_y)
\end{eqnarray*}
with  $t_1=0.24, t_2 =0.52, t_2' = -0.1, t_{s,d} = (t_2 \pm t_2')/2, t_{3s} = t_{3d} \sim 0, \mu = -0.273$. The matrix elements and electron operators ($d_{1,k}$ and $d_{2,k}$) for the other copy with the $y'z$ orbital on the $A$ and $x'z$ on the $B$ sublattice can be obtained by performing the $S_4$ symmetry transformation as demonstrated in \cite{Hu2012-S4}.

\begin{figure}[h!]
\caption{\label{ElectronOverdoped}Raman vertices and non-resonant responses for the electron overdoped chalcogenide ($\mu = -0.273, t_2'  = -0.2 $ all units in $eV$). a) Response in the superconducting state as observed in \cite{Kretzschmar2013}. b) Response evaluated theoretically using the model described in the text. c) Square of the Raman weight factors on the Fermi surface in the three different geometries. We have used a $\Delta_0 cos k_x cos k_y$ form factor for the gap with $\Delta_0$ = 0.1. The laser frequency of $\sim 2.3 eV$ corresponding to optical frequencies used in Raman experiments is chosen. The results are robust to changes in the laser frequency as long as it is away from resonance. An impurity broadening of $5 meV$ is chosen. The notation of our irreps corresponds to an axis rotated by $\pi/4$ with respect to that used in ref \cite{Kretzschmar2013}.  }
\includegraphics[width=0.5\textwidth]{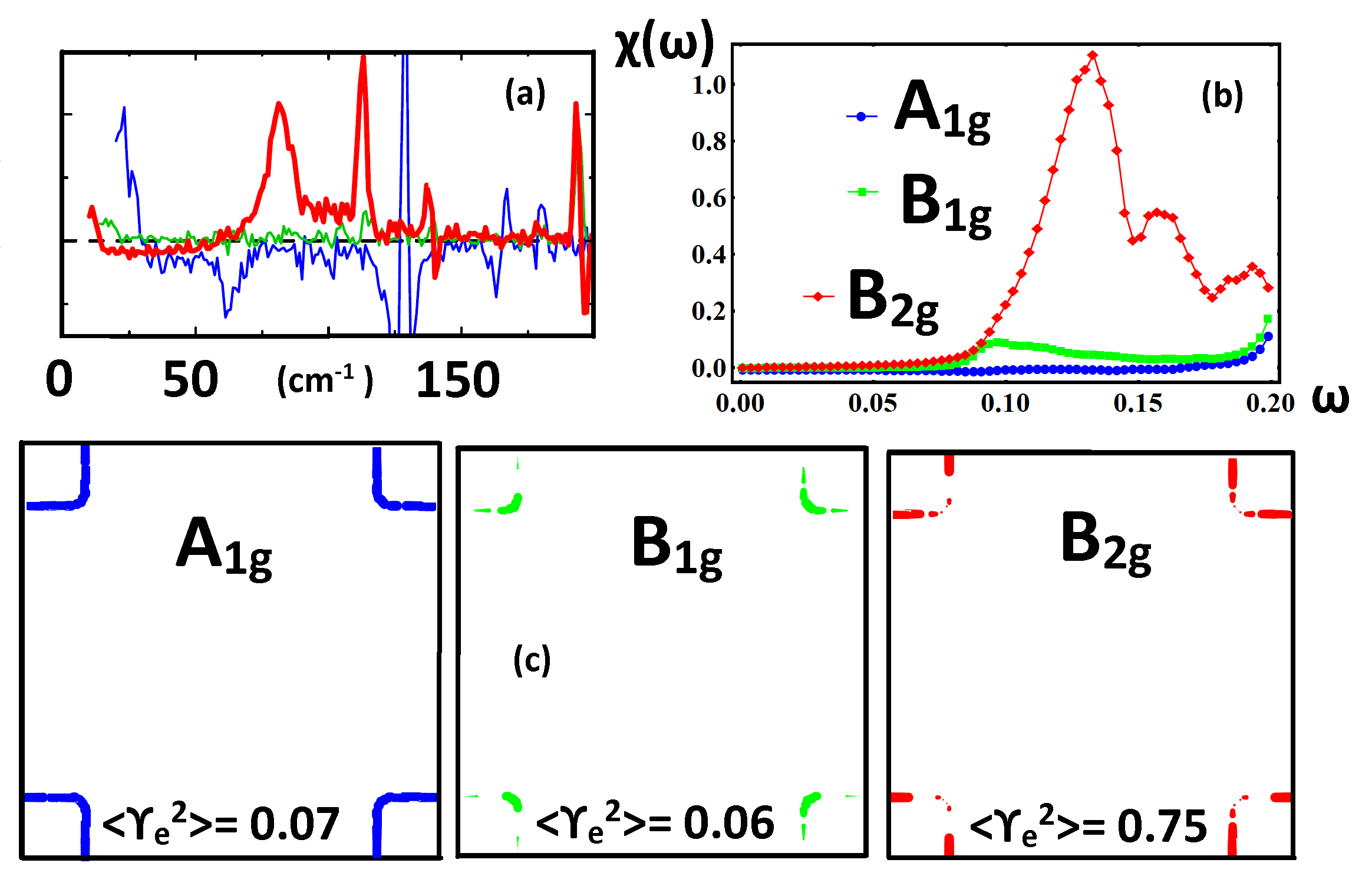}
\end{figure}

\begin{figure}[h!]
\caption{\label{ElectronLightdoped}Same as the caption in fig \ref{ElectronOverdoped} but for the case of electron doped pnictide $Ba(Fe_{1-x}Co_x)_2As_2$, ($\mu = -0.273, t_2' = -0.17 $ in $eV$). The gap on the electron pocket is chosen to be smaller (by $\sim 25 $ percent) than that of the hole pocket consistent with experiment \cite{Takahashi2009}. Data in $(a)$ and $(b)$ taken from \cite{Muschler2009}. Note that the vertex on the hole pockets in the $B_{1g}$ and $B_{2g}$ geometries is negligible to be seen. The labelling of the irreps is similar to that used in \cite{Muschler2009}.  }
\includegraphics[width=0.5\textwidth]{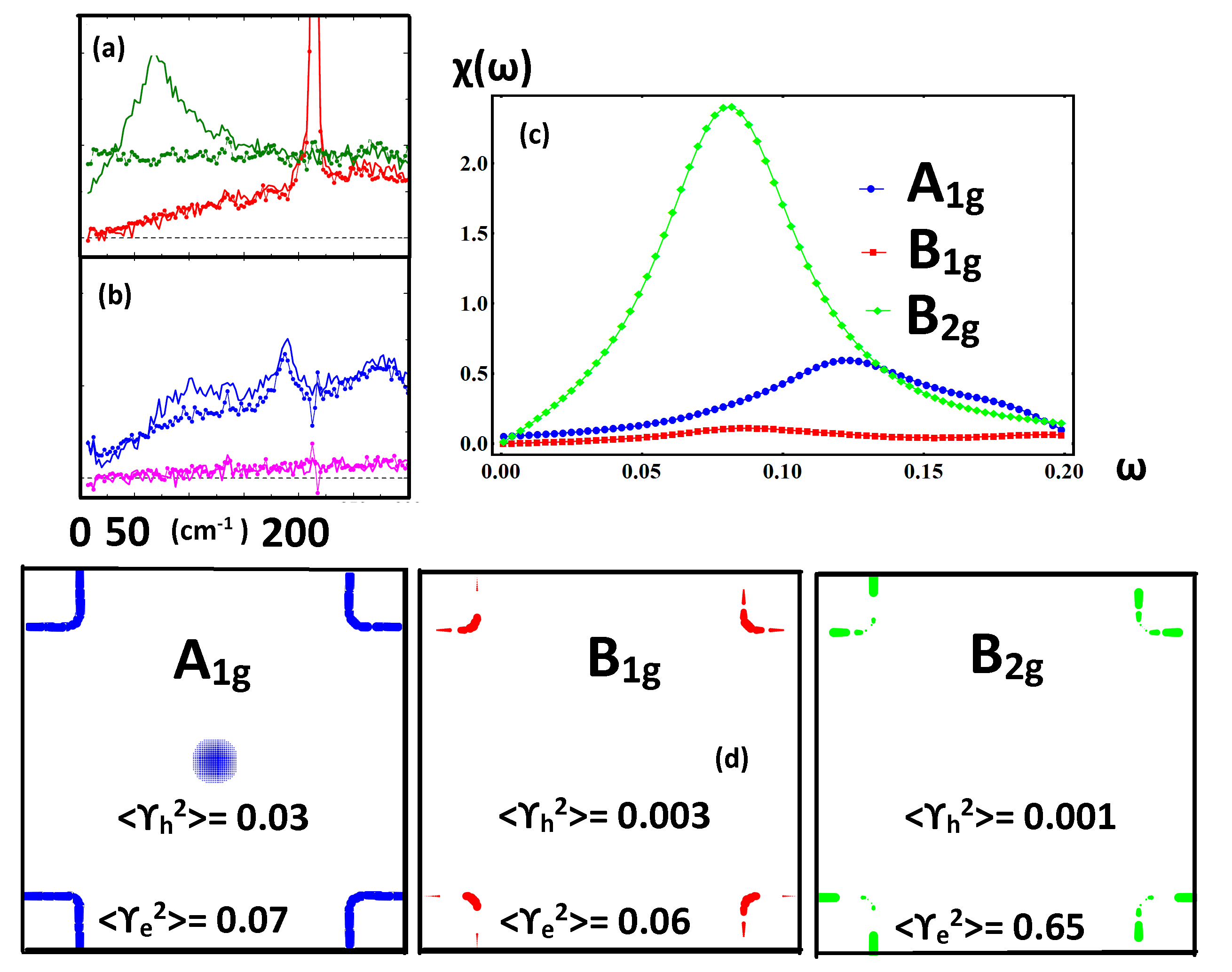}
\end{figure}

\begin{figure}[h!]
\caption{\label{Holedoped}Same as the caption in fig \ref{ElectronOverdoped} but for the case of hole doped pnictide $Ba_{0.6}K_{0.4}Fe_2As_2$ ($\mu = -0.21, t_2' = -0.1 $ in $eV$). The gap on the inner hole pocket is chosen to be anisotropic by 10 percent to fit experimental data. The overall intensity in the $B_{1g}$ geometry is multiplied by $\approx 0.8$ to match the data. Data in $(a)$ taken from \cite{Kretzschmar2013}. As before, our irreps are labelled according to an axis rotated by $\pi/4$ with respect to that used in ref \cite{Kretzschmar2013}.}
\includegraphics[width=0.55\textwidth]{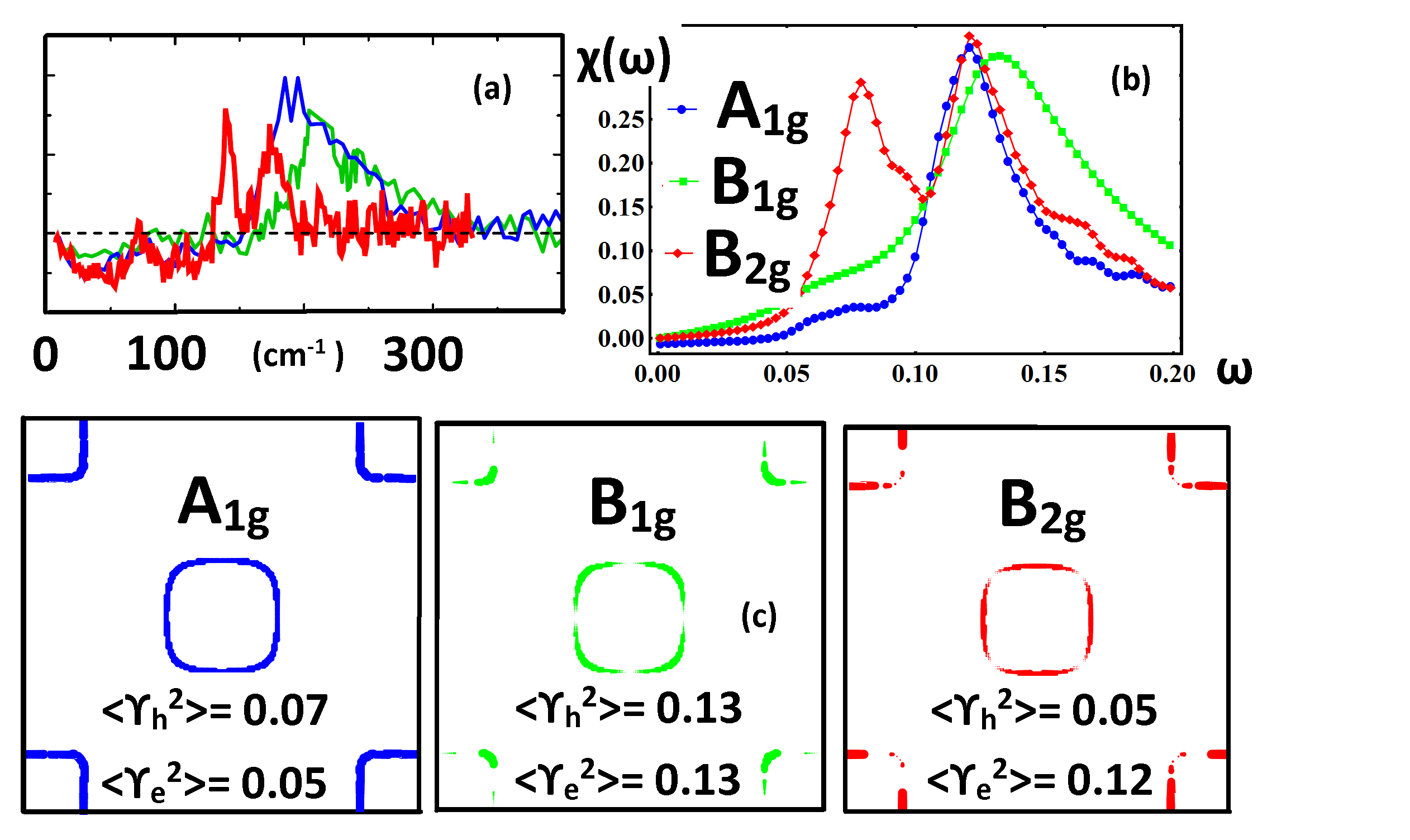}
\end{figure}

We now evaluate and compare the Raman responses for three different cases of the 122-type Iron based superconductors as reported in \cite{Kretzschmar2013,Muschler2009}.
\\
\\
\textit{ Electron overdoped chalcogenide} ($Rb_{0.8}Fe_{1.6}Se_2$) : We begin with the simplest case of the electron overdoped Chalcogenides with hole pockets absent from the $\Gamma$ point. The Fermi surfaces observed in ARPES \cite{Ding2011-122Chal,Zhou2011-PRL,Zhou2011-PRB}  for the electron overdoped chalcogenide consist of two electron pockets at ($\pm \pi, \pm \pi$) points in the reduced Brillouin zone.
Fig \ref{ElectronOverdoped} (c) shows   the plots of the  Fermi surfaces seen  with different polarization geometries. The values shown near the Fermi surface denote the average value of the square of the Raman vertex in each geometry, on the fermi surface. The $A_{1g}$ case sees an isotropic weight all across the fermi surface whereas the $B_{1g}$ and $B_{2g}$ geometries sample two different regions on the electron pockets (A larger color intensity corresponds to a greater value of the weight. Intensity plots shown are relative and cannot be compared between two geometries). The Raman response for the different polarization geometries is plotted in fig \ref{ElectronOverdoped} (b). The response in the $B_{2g}$ geometry (with respect to the two atom unit cell) completely dominates when compared to the $A_{1g}$ and $B_{1g}$ cases. This is because of the large weight present on the electron pocket which compensates for its small density of states. The small value of the vertex, in combination with the small density of states present on the electron pocket, leave the responses in the $A_{1g}$ and $B_{1g}$ geometries negligible in comparison with the $B_{2g}$ channel. 
\newline
\newline
\textit{ Electron doped pnictide} ($Ba(Fe_{1-x}Co_{x})_2 As_2$): In this case, ARPES \cite{Takahashi2009} sees a small hole pocket centered around the $\Gamma$ point with the usual two electron pockets at the ($\pm \pi, \pm \pi$) points in the reduced brillouin zone (See fig \ref{ElectronLightdoped} (d) ). The distribution of the Raman weight on the electron pockets is very similar to that of the electron overdoped chalcogenide discussed previously. However, on the hole pocket, the weight in the $A_{1g}$ geometry is about an order of magnitude larger than that in the other two geometries. This, combined with the large density of states at the hole pocket  which is captured in the $S_4$ model, provides  a low intensity broad peak at high energies which  consists of contributions from the hole pocket and the mixed response term in eq. \ref{eq:ImResponse}. The reason for the high intensity peak in the $B_{2g}$ geometry is similar to the previous case because of the large Raman weight on the electron pockets.
\newline
\newline
 \textit{Hole doped pnictide}($Ba_{0.6}K_{0.4}Fe_2As_2$): Fig \ref{Holedoped} shows the Fermi surface, vertex values and the response for the hole doped Iron-pnictides. The experimental fermi surface \cite{Borisenko2009-122,Ding2008-122,Zhou2008-122} consists of two large hole pockets centered around the $\Gamma$ point with electron pockets at the ($\pm \pi, \pm \pi$) points in the reduced Brillouin zone. The distribution of the Raman weight is different compared to the previous two cases(fig \ref{Holedoped} (c) ). In the $A_{1g}$ and $B_{1g}$ geometries, the hole and the electron pockets have comparable weights. However, due to its large density of states, the intensity from the hole pocket completely dominates the spectrum. In contrast, in the $B_{2g}$ case, the hole pocket has a smaller value of the vertex compared to the electron pocket. However, the large density of states compensates for the small value of vertex so that the final response has almost equal contributions from the hole pocket and the electron pocket, giving rise to two distinct peaks at different energies. If, however, according to our picture, the smaller and larger energy peaks in the $B_{2g}$  emerge from the electron and hole pockets respectively, the experimentally observed energy difference between the two shouldn't be more than $\sim 15-20 cm^{-1}$.  However we see that the energy difference between the two peaks is observed to be $ \sim 35 cm^{-1}$ ( see for example \cite{Takahashi2009-EPL} ), a few ten $cm^{-1}$ more. In both  $B_{2g}$ and $A_{1g}$ geometries, there are also  low energy features (around 80 $cm^{-1}$) in experimental measurements  as shown in Fig3. (a). These features stem from the outer hole pocket that is not present in the $S_4$ model because the $S_4$ model only provides two hole pockets rather than three hole pockets at $\Gamma$ point.
Both ARPES \cite{Borisenko2009-122,Ding2008-122,Zhou2008-122} and Raman \cite{Kretzschmar2013} have shown a much smaller  gap on the outer hole pocket than those on the electron pockets( the ratio between the two gaps is about $\sim 0.4$).  We note that in all above calculations,  the gap values that we have chosen for all the three materials is consistent with experiments.   
\newline
\newline
From the above results, we can see that the detailed structure  of  the Raman vertex and the band curvature are essential in determining the overall intensity of the Raman spectrum.  In literature, the Raman vertex is often obtained by using the effective mass approximation; here however,  it is determined exactly by taking into account of the non-zero laser frequency. In the case of Iron superconductors, this makes a significant difference - the degeneracy at the $M$ point (in the folded BZ) would have overestimated the vertex and the intensity. Specifically, a naive use of such an approximation would mean that (i) in the $Co$ doped 122 case, the $B_{2g}$ peak would be a lot more exaggerated and (ii)  in the hole doped 122 case, the peaks arising from the hole pockets  would be heavily overshadowed by the contribution from the electron pockets. \\
\newline
 For the electron doped chalcogenide and the electron doped pnictide compounds, our result gives a reasonable quantitative description of the origin of the peaks, their frequencies and intensities, which was only qualitatively argued in refs. \cite{Kretzschmar2013} and \cite{Muschler2009}. In the hole doped 122 case, apart from the more quantitative picture, there are  several significant qualitative differences  in our interpretation of the data. Firstly, in our calculation, the two large peaks observed in the $A_{1g}$ and $B_{1g}$ (2- $Fe$ unit cell notation) in the hole doped 122 compound are from the central hole pocket in contrast to \cite{Kretzschmar2013}, where the authors argued that the peaks originate from the electron pocket by ruling out the hole pocket as a possible cause (supplemental material from \cite{Kretzschmar2013}). Their essential argument is that  there is an energy difference about  ($\sim 15 cm^{-1}$)  between the peak positions measured in the two cases. However such a small energy difference can be easily explained by the gap variations in  the hole pockets\cite{Borisenko2009-122, Takahashi2009-EPL} instead of ruling out hole pocket contributions.  Secondly, our calculation shows that the two large peaks observed in the $B_{2g}$ geometry originate from both electron and hole pockets.   The existence of either the Bardasis-Schrieffer \cite{Kretzschmar2013} or orbital resonance \cite{Phillips2013} modes suggested in the previous  studies, is not necessary.  Finally,    the previous conclusion in \cite{Kretzschmar2013}  about the existence of  the large gap anisotropy on the $\delta$ (outer) electron pocket in the hole doped 122 case is not reliable.  The conclusion - drawn by observing a combination of the small shoulders  close to 80 $cm^{-1}$ (2 $\Delta_{min}$) in the $A_{1g}$ and $B_{2g}$ geometries along with the higher energy peaks close to 200 $cm^{-1}$ (2 $\Delta_{max}$) - is only backed by analogous features seen in a calculation performed in \cite{Scalapino2009-Raman}. However,  such an interpretation is only valid if the hole pockets do not make a large contribution while our calculation clearly shows that  the hole pocket contribution is significant. More importantly, this previous interpretation is incompatible  with ARPES results  that shows an almost isotropic  gap on electron pockets.

In conclusion, we have shown that the two weakly coupled single orbital models proposed in \cite{Hu2012-S4} successfully describe the Raman spectra in the superconducting state of the hole doped and electron doped pnictide and chalcogenide superconductors. This is done with a simple $cos k_x cos k_y$ form of the superconducting gap in the Brillouin zone without invoking additional competing forms of the gap.The results are in agreement with the experimental data.

We thank discussion with H. Ding and NN Hao.  The work is supported by the Ministry of Science and Technology of China 973
program(2012CB821400), NSFC, and   the Strategic Priority Research Program of  CAS (Grant No. XDB07000000).

\bibliographystyle{apsrev}
\bibliography{Raman}

\end{document}